# On the validity of the classical apsidal motion formula for tidal distortion

Eliot J. Quataert, Pawan Kumar[1] and Chi On Ao

Department of Physics

Massachusetts Institute of Technology, Cambridge, MA 02139

**Abstract:** We check the validity of the widely used classical apsidal motion formula as a function of orbital parameters, stellar structure, and stellar rotation rate by comparing dynamical calculations of the periastron advance with the static tidal formula. We find that the classical formula gives very accurate results when the periods of the low order quadrupole g, f and p modes are smaller than the periastron passage time by a factor of about 7 or more. However, when this condition is not satisfied, the difference between the classical formula and the exact result can be quite large, and even periastron recession can result. The largest difference arises when one of the low order modes of the star is nearly resonant with an integer multiple of the orbital frequency minus twice the rotation rate of the star. The resonance of higher order g-modes (number of radial nodes $\gtrsim 4$) with the orbit is very unlikely to cause significant deviation from the classical result because of their weak coupling to the tidal force and thus their small contribution to the apsidal motion. Resonances involving rotational modes of the star are also unlikely to make much contribution to the apsidal motion because of their small overlap with the tidal force, even though they have periods comparable to the periastron passage time.

We apply our work to two famous binary systems (AS Cam and DI Her) which show abnormally small apsidal motion, and conclude that dynamical effects are unimportant for these systems, i.e. the static tide assumption is an excellent approximation.

---

# 1. Introduction

When the gravitational field of a star differs from that of a Newtonian point mass, the orbit of its companion will deviate from a Keplerian orbit. To lowest order, a perturbation to the $1/r$ gravitational potential causes the periastron to rotate. This is the origin of apsidal motion. There are primarily three effects which cause deviation from the $1/r$ gravitational potential: the General Relativistic correction to Newtonian gravity, the quadrupole moment that arises due to the rotational distortion of a star, and the quadrupole moment due to tidal distortion. The first two effects are relatively easy to calculate and are well understood. The third effect, the modification of the gravitational potential due to tidal distortion, displays more complex behavior and is the subject of this paper. The derivation of the classical apsidal motion formula, due to tidal distortion, was first worked out by Cowling (1938) and Sterne (1939) and is widely used in the interpretation of observed apsidal motion. Their derivation follows from the approximation that the star can instantaneously adjust its equilibrium shape to the time-dependent tidal force as it orbits around the center of mass of the system. This approximation is clearly valid in the limit that the periods of the low order modes of oscillation of the star are much smaller than the orbital period (for orbits of low eccentricity) or the periastron passage time (for eccentric orbits). However, this approximation is expected to fail when the mode periods are of order the relevant time scale. Indeed, as Papaloizou and Pringle (1978) have pointed out, resonances can significantly modify the classical formula.

In view of the wide use of the classical apsidal formula, it is important to understand where the classical approximation is valid and where it fails. In this paper, we model the tide dynamically as a linear superposition of the non-radial oscillations of a star excited by the gravitational tidal force of the primary. We calculate, using a fourier series expansion of the oscillation amplitude, the quadrupole gravitational potential perturbation due to these oscillations, and the resulting periastron shift. Our result is compared with the classical apsidal motion formula for stars of different polytropic index and mass, and for various orbital parameters. The rotation of the star and its modification of resonances is also considered. We also consider the effects of rotational modes on apsidal motion. However the modification of p and g mode eigenfunctions due to stellar rotation, which can be quite substantial, is not included in our calculations.

The plan of our paper is as follows. In §2 we present the basic equations for the perturbation to the gravitational potential arising from dynamic tides and discuss the validity of the classical apsidal formula for nearly circular orbits. In §3, we examine the effects of tidal resonances on apsidal motion as a function of modal and orbital parameters. In §4, we compare our results for polytropic stars with the classical apsidal formula to find the system parameters for which deviation from the classical prediction is expected. The effect of rotational modes on apsidal motion is considered in §5. In §6 we examine the apsidal motion of two peculiar binary systems (AS Cam and DI Her) which have observed periastron advances that differ significantly from the predicted values. We investigate whether the discrepancy might in fact be due to the inapplicability of the classical apsidal formula. Our results are summarized and discussed in §7.

## 2. Basic Equations for tidal perturbation and apsidal motion for nearly circular orbits

Consider a binary system where one of the stars, taken to be a point object, has mass $M$ and its companion star, the secondary, has mass $M_*$. The position of the primary star in the orbit, as seen by the secondary star, is specified by $R(t)$, the separation between them, and the angular position, $\phi_o(t)$. The secondary is assumed to be rotating as a solid body with angular speed $\Omega_*$ about an axis perpendicular to the orbital plane (taken to be the z-axis).

The perturbation to the density of the secondary due to the tidal gravitational field of the primary can be written as a sum over its normal mode amplitudes:

$$\delta\rho(\mathbf{r},t) = \sum_{n\ell m} A_{n\ell m}(t)\delta\rho_{n\ell}(r)Y_{\ell m}(\theta,\phi), \qquad (1)$$

where $\delta\rho_{n\ell}$ is the normalized density eigenfunction, $Y_{\ell m}(\theta,\phi)$ is the spherical harmonic function, and $A_{n\ell m}$ is the mode amplitude which is calculated from the tidal gravity of the primary as described below. The pressure (p) and gravity (g) modes are uniquely specified by three numbers: $n$, the number of nodes in the radial direction, the spherical harmonic degree, $\ell$, and the azimuthal number, $m$. We use a short hand notation, $\alpha$, to denote the collective index $(n,\ell,m)$. For conciseness of notation, we take $n$ to be positive for p-modes and negative for g-modes.

The gravitational potential, at a point $\mathbf{R}(t)$, due to the tidally perturbed star is:

$$\Psi(\mathbf{R}, t) = -G \int d\mathbf{r} \frac{\rho(\mathbf{r})}{|\mathbf{R} - \mathbf{r}|}. \tag{2}$$

Substituting the normal mode expansion of the density from equation (1) and using the standard spherical harmonic expansion of $1/(|\mathbf{R} - \mathbf{r}|)$ (Jackson, 1962), we obtain the following expression for the perturbation to the gravitational potential:

$$\Psi_1(\mathbf{R}, t) = -4\pi G \sum_{nlm} \frac{A_{nlm} Y_{lm}(\theta_o = \pi/2, \phi_o(t) - \Omega_* t) Q_{nl}}{(2l+1) R(t)^{l+1}}, \tag{3}$$

where the orbital plane is taken to be $\theta_o = \pi/2$,

$$Q_{nl} \equiv \int_0^{R_*} dr \, r^{\ell+2} \delta\rho_\alpha(r) \tag{4}$$

is the overlap integral for the mode, $R_*$ is the radius of the star, and the displacement eigenfunctions, $\boldsymbol{\xi}_{n\ell m}$, are normalized to have unit energy, i.e.,

$$\omega_\alpha^2 \int d^3x \, \rho \boldsymbol{\xi}_{n\ell m} \cdot \boldsymbol{\xi}_{n\ell m} = 1. \tag{5}$$

The equation for the mode amplitude, $A_\alpha$, can be cast in the form of the following second order ordinary differential equation for a forced, damped harmonic oscillator:

$$\frac{d^2 A_\alpha}{dt^2} + 2\Gamma_\alpha \frac{dA_\alpha}{dt} + \omega_\alpha^2 A_\alpha = f_\alpha(t), \tag{6}$$

with

$$f_\alpha(t) = \frac{4\pi}{(2\ell+1)} \frac{GM\omega_\alpha^2 Q_{nl}}{R^{\ell+1}(t)} Y_{\ell m}^*(\theta_o = \pi/2, \phi_o(t) - \Omega_* t) \equiv \tilde{f}_{\ell m} \frac{4\pi GM\omega_\alpha^2 Q_{nl}}{(2\ell+1) a^{\ell+1}}, \tag{7}$$

as derived for a nonrotating star by Press and Teukolsky (1977), where $\Gamma_\alpha$ is the dissipation rate of the mode, $a$ is the semimajor axis of the orbit, and where the reduced forcing function, $\tilde{f}_{\ell m}(t) = Y_{\ell m}^*(\pi/2, \phi_o(t) - \Omega_* t)/[R(t)/a]^{\ell+1}$, contains all the time dependence of $f_\alpha(t)$. The amplitude is as seen from a frame rotating with the star. The rotating coordinate system is related to the non-rotating coordinate system through a rotation about the z-axis by an angle of $\Omega_* t$.

The solution of equation (6), in terms of the Green's function, is

$$\tilde{A}_\alpha(t) \equiv A_\alpha(t)\frac{(2\ell+1)a^{\ell+1}}{4\pi GM\omega_\alpha^2 Q_{nl}} = \frac{\Omega_o^2}{\omega_\alpha'}\int_{-\infty}^t dt_1 \exp\left[-\Gamma_\alpha(t-t_1)\right]\sin\left[\omega_\alpha'(t-t_1)\right]\tilde{f}_{\ell m}(t), \quad (8)$$

where $\omega_\alpha' \equiv \sqrt{\omega_\alpha^2 - \Gamma_\alpha^2} \approx \omega_\alpha$ and where $\tilde{A}_\alpha(t)$, the reduced mode amplitude, contains all the time dependence of $A_\alpha(t)$.

We next derive an expression for $\Psi_1$ and the periastron advance for nearly circular orbits, and then present results for orbits of arbitrary eccentricity. For nearly circular orbits $\phi_o(t) \approx \Omega_o t$, where $\Omega_o$ is the angular velocity of the star, and $R(t)$ is approximately constant. With these two simplifications the integral in equation (8) is straightforward to carry out and the resultant mode amplitude is given by:

$$\tilde{A}_\alpha(t) \approx \frac{\Omega_o^2 Y_{\ell m}(\pi/2,0)\exp(-im\Omega_1 t)}{\omega_\alpha^2 - m^2\Omega_1^2 - i2m\Gamma_\alpha\Omega_1}, \quad (9)$$

where

$$\Omega_1 = \Omega_o - \Omega_* \quad (10)$$

Inserting this into equation (3) and performing the sum over $m=0$ and $m=\pm 2$ for the dominant $\ell=2$ modes, we find for the perturbation to the gravitational potential,

$$\Psi_1 \approx \frac{-\pi G^2 M}{5R^6(t)}\sum_n Q_{n2}^2\left[1 + \frac{3(1 - 4\Omega_1^2/\omega_\alpha^2)}{(1 - 4\Omega_1^2/\omega_\alpha^2)^2 + 16\Gamma_\alpha^2\Omega_1^2/\omega_\alpha^4}\right]. \quad (11)$$

We have ignored the splitting of the eigenfrequencies due to rotation in the above equation. The perturbation to the gravitational potential of a star due to tides, in the static tidal approximation, is given by (Schwarzchild, 1958):

$$\Psi_1 = \frac{-2kR_*^5 GM}{R^6}, \quad (12)$$

where $k$, the dimensionless apsidal motion constant, depends on the structure of the star. Comparison of equations (11) and (12) leads to the following expression for the apsidal constant $k$,

$$k = \frac{\pi G}{10R_*^5}\sum_n Q_{n2}^2\left[1 + \frac{3(1 - 4\Omega_1^2/\omega_\alpha^2)}{(1 - 4\Omega_1^2/\omega_\alpha^2)^2 + 16\Gamma_\alpha^2\Omega_1^2/\omega_\alpha^4}\right] \quad (13)$$

In the limit that $|\Omega_1| \ll \omega_\alpha$ for the low order modes, which have the largest values of the overlap integral, the star can adjust its shape to the instantaneous tidal force, *i.e.*, the tide is static, and the above expression for $k$ reduces to

$$k \approx \frac{2\pi G}{5R_*^5} \sum_n Q_{n2}^2. \tag{14}$$

The classical apsidal formula uses the value of $k$ given by the above expression. Thus, we expect deviation from the classical formula (see eq. [13]) when the frequency of one of the low order modes is comparable to $2\Omega_1$. This result, as we show in the next section, applies to all orbits (not just nearly circular ones), although for eccentric orbits resonances with harmonics of the orbital period must also be considered. We also note that the periastron advance is linearly related to the apsidal motion constant and thus our correction to the apsidal motion constant is also a linear correction to the periastron advance.

### 3. The effect of resonance on the apsidal motion constant

In this section we consider the effect of orbital resonance on a given mode's contribution to the periastron advance. In the next section (§4), we consider whether or not this resonance has an effect on the apsidal constant for a polytropic star by considering its complete normal mode spectrum. The two parameters defined below are of central importance in determining when the classical apsidal motion formula is valid:

$$b_\alpha \equiv \frac{\omega_\alpha}{\Omega_p}, \qquad r_\alpha \equiv \frac{\omega_\alpha}{\Omega_o} = \omega_\alpha \sqrt{\frac{a^3}{GM_t}}, \tag{15}$$

where $a$ is the semi-major axis, $M_t$ is the total mass of the system, $\Omega_p$, the angular velocity at periastron, is given by:

$$\Omega_p \equiv \Omega_o \frac{\sqrt{1-e^2}}{(1-e)^2}, \tag{16}$$

and $e$ is the orbital eccentricity. For a nearly circular orbit, the two parameters, $b_\alpha$ and $r_\alpha$, are approximately the same, but for eccentric orbits they are quite different. As we show below, the apsidal motion can be quite different from that given by the static tide approximation when $b_\alpha$ is small ($b_\alpha \lesssim 7$).

To start our analysis, we separate out the $\Omega_*$ dependent term in the reduced forcing function, $\tilde{f}_{\ell m}(t, \Omega_*) = \exp(im\Omega_* t)\tilde{f}_{\ell m}(t, \Omega_* = 0)$. Without rotation, the forcing function is periodic with period $T$ and we can decompose it in terms of its Fourier coefficients. Therefore,

$$\tilde{f}_{\ell m}(t, \Omega_*) = \exp(im\Omega_* t) \left[ \sum_{n=1}^{\infty} C_n^{(\ell m)} \sin(n\Omega_o t) + \sum_{n=0}^{\infty} D_n^{(\ell m)} \cos(n\Omega_o t) \right]. \quad (17)$$

It is straightforward to show that $Im(D_n^{(\ell m)}) = C_n^{(\ell m)} = 0$ for $m=0$ modes. In addition, for $n \gtrsim \Omega_p/\Omega_0$, $Re(D_n^{(2m)}) \approx -\text{sign}(m)Im(C_n^{(2m)})$ for $m = \pm 2$ modes. The $m = 0$ Fourier coefficients decrease monotonically with $n$, whereas the $m = \pm 2$ coefficients have a maximum at $n \approx 2\Omega_p/\Omega_o$ and are larger in magnitude than the $m = 0$ coefficients for $n \gtrsim \Omega_p/\Omega_o$. For $n$ greater than about $5\Omega_p/\Omega_o$, the $D_n^{(20)}$ fall off as $\exp(-1.3ne^{-0.25}\Omega_o/\Omega_p)/(1-e)^{(3/2)}$ and the $D_n^{(22)}$ are larger than the $D_n^{(20)}$ by about a factor of $\exp(2.6e^{-.25})$.

Inserting the Fourier series expansion of $\tilde{f}_{\ell m}$ (eq. [17]) into equation (8) and carrying out the integral, the amplitude for mode $\alpha$, in the rotating reference frame, can be written in terms of the above Fourier coefficients as follows:

$$\tilde{A}_\alpha(t) = \frac{\exp(im\Omega_* t)}{2} \left[ \sum_{n=1}^{\infty} \frac{(D_n^{(\ell m)} + iC_n^{(\ell m)}) \exp(-in\Omega_0 t - i\phi_n)}{\sqrt{[r_\alpha^2 - (n-ms)^2]^2 + 4d_\alpha^2(n-ms)^2}} \right.$$
$$\left. + \sum_{n=1}^{\infty} \frac{(D_n^{(\ell m)} - iC_n^{(\ell m)}) \exp(in\Omega_0 t - i\phi'_n)}{\sqrt{[r_\alpha^2 - (n+ms)^2]^2 + 4d_\alpha^2(n+ms)^2}} + \frac{2D_0^{(\ell m)}}{r_\alpha^2 - m^2s^2 + 2imsd_\alpha} \right] \quad (18)$$

where

$$d_\alpha = \frac{\Gamma_\alpha}{\Omega_o}, \quad s = \frac{\Omega_*}{\Omega_o}, \quad \tan\phi_n = \frac{2d_\alpha(ms-n)}{r_\alpha^2 - (n-ms)^2}, \quad \tan\phi'_n = \frac{2d_\alpha(ms+n)}{r_\alpha^2 - (n+ms)^2} \quad (19)$$

It is clear from equation (18) that $m=0$ mode amplitudes are unaffected by rotation (provided that we ignore the perturbation to the mode's eigenfuntion) and are resonantly excited when $r_\alpha$ is an integer. For nonzero $m$, resonance occurs when $r_\alpha = |n \pm ms|$. The magnitude of the resonant term in the series, ignoring for the moment the effect of rotation, is approximately $C_{\lfloor r_\alpha \rfloor}^{(\ell m)} \delta r_\alpha^{-1}$, where $\lfloor r_\alpha \rfloor$ is the nearest integer to $r_\alpha$ and $\delta r_\alpha \equiv r_\alpha - \lfloor r_\alpha \rfloor$. Similarly, the magnitude of the $n$-th non-resonant term in the series is approximately $C_n^{(\ell m)}/r_\alpha$. Thus, the contribution of the resonant term to the mode amplitude is of the same order as the sum of the non-resonant terms provided that

$$\sum_{n=1}^{\lfloor r_\alpha \rfloor - 1} \frac{C_n^{(\ell m)}}{r_\alpha} \approx \frac{\exp(-1.3b_\alpha e^{-.25})}{(1-e)^{(3/2)}\delta r_\alpha}. \quad (20)$$

Using the expression for $C_n^{(lm)}$ given above, we find that the resonant term is of order the sum of the non-resonant terms provided that $b_\alpha \lesssim -e^{+0.25} \ln(\delta r_\alpha)$. The evolution of the orbit limits the value of $\delta r_\alpha$ to be greater than some minimum value, $\delta r_{min}$, which is of order $10^{-3}$ for close binaries; $\delta r_\alpha \lesssim \delta r_{min}$ is unphysical because changes in the orbital period, as a result of orbital evolution, cause the phase of the mode at periastron to drift by $\pi$ in $\delta r_{min}^{-1}$ orbits (see Kumar, Ao, & Quataert, 1995). Thus, resonances have a significant effect on the mode amplitude, the perturbation to the gravitational potential, and the apsidal constant only when $b_\alpha$ is small (less than about 7 for $\delta r_{min} \approx 10^{-3}$).

For our subsequent discussion of the validity of the classical apsidal motion formula, it is useful to define the following dimensionless number, $\zeta_n$, which gives the fractional difference between the gravitational potential, due to a mode of order $n$, in the static tide limit and in the exact calculation:

$$\zeta_n = \frac{\Psi_{1,n} - \Psi_{1,n}^s}{\sum_i \Psi_{1,i}^s}, \tag{21}$$

where $\Psi_{1,n}$ is the perturbation to the gravitational potential arising from the $n$-th mode and $\Psi_{1,n}^s$ is the perturbation to the gravitational potential in the static tide limit (obtained by setting $A_\alpha = f_\alpha/\omega_\alpha^2$ in eq. [3]). Using equations (3) and (8) to determine $\Psi_{1,n}$ and equations (12) and (14) to determine $\Psi_{1,n}^s$, we obtain:

$$\zeta_n = \frac{Q_{n2}^2}{\sum_i Q_{i2}^2} \left[ \frac{4\pi R_{peri}^3 r_\alpha^2}{5a^3} \sum_{m=0,\pm 2} \tilde{A}_{n2m}(t=0) Y_{2m}(\theta_0 = \pi/2, \phi_0 = 0) - 1 \right], \tag{22}$$

where $t = 0$ corresponds to the periastron and $R_{peri} = a(1-e)$ is the periastron distance. By definition, $\zeta_n = 0$ in the static tide limit, whereas $\zeta_n = 0.5$, for example, would mean a 50% deviation from the classical formula because of a resonance involving a quadropole mode of order $n$.

Since the tidal potential falls off as $1/R^6$, the periastron advance is primarily determined by the tidal potential at periastron. That is why we have chosen the mode amplitude at $t = 0$ (periastron) in equation (22). We note, however, that $\zeta_n$ is only an approximate characterization of the deviation from the classical apsidal motion formula. The correct and accurate description of the deviation from the classical apsidal formula is provided by explicitly calculating the periastron advance using the perturbation to the gravitational potential, $\Psi_1$ (eq. [3]), and comparing the result with the classical formula. We thus introduce a parameter $\eta_n$, which gives the fractional deviation, arising from the $n$-th mode, from the total periastron advance given by the classical apsidal motion formula.

We find from numerical calculations that the functional dependence of $\zeta_n$ on the modal and orbital parameters is very similar to that of $\eta_n$. Figure (1a) shows the comparison between $\zeta_n$ and $\eta_n$ as a function of $b_\alpha$. The stronger $b_\alpha$ dependence of $\eta_n$ is due to the stronger $b_\alpha$ dependence of the mode amplitude away from periastron (Kumar, Ao, & Quataert, 1995). The discrepancy between $\eta_n$ and $\zeta_n$ is most significant for $b_\alpha \lesssim 1$. Since we are primarily concerned with deviation from the classical apsidal formula in the neighborhood of a resonance, figure (1b) shows the functional dependance of $\zeta_n$ and $\eta_n$ on $\delta r_\alpha$. Because $\zeta_n$ and $\eta_n$ are roughly equal, as illustrated in figure (1), we present our subsequent analysis of the validity of the classical apsidal motion formula in terms of the more analytically tractable quantity, $\zeta_n$.

In this section, as we are only considering individual modes and their contribution to the apsidal motion formula, we take $Q_\beta = \delta_{\alpha,\beta} Q_\alpha$ or $Q_\alpha^2 / \sum_\beta Q_\beta^2 = 1$ and investigate in some detail the effect of a resonance on $\zeta_n$ as a function of orbital parameters. This drastic simplification of the normal mode spectrum to a single mode is made to gain physical insight. In the next section we carry out the consistent calculation and include the entire modal spectrum of polytropic stars. We now substitute the Fourier series expansion of the mode amplitude (eq. [18]) into equation (22). We note, however, that the Fourier series technique employed in this paper is only applicable so long as the orbital evolution time scale is such that the resulting change in $r_\alpha$ per orbit is less than $(\delta r_\alpha)^2$ (this condition is obtained by requiring that the tidal forcing of a mode is in phase at periastron over $1/\delta r_\alpha$ orbits, in order for the the mode amplitude to build up). Keeping only the $\ell = 2$ terms in equation (22), we obtain

$$\zeta_n = \sqrt{\frac{\pi}{5}} \frac{R_{peri}^3}{a^3} r_\alpha^2 \sum_{n=1}^{\infty} \sum_{j=1,3} \Delta_j \frac{(r_\alpha^2 - n_j^2)}{(r_\alpha^2 - n_j^2)^2 + 4 d_\alpha^2 n_j^2} - 1, \qquad (23)$$

where

$$\Delta_1 = \sqrt{\frac{3}{2}} (D_n^{(22)} - i C_n^{(22)}), \quad \Delta_2 = \sqrt{\frac{3}{2}} (D_n^{(22)} + i C_n^{(22)}), \quad \Delta_3 = -D_n^{(20)}, \qquad (24)$$

and

$$n_1 = n + 2s, \quad n_2 = n - 2s, \quad n_3 = n. \qquad (25)$$

The sum over $j$ in equation (23) corresponds to the sum over $m = 0, \pm 2$. Note that we have again ignored the rotational splitting of the eigenfrequencies in the above derivation.

For most physical cases of interest, we can accurately take $Re(D_n^{(22)}) \approx -Im(C_n^{(22)})$ and thus $\Delta_1 \approx 0$. We see from equation (23) that resonance occurs when $r_\alpha = |n - 2s|$ or when $r_\alpha = n$, causing the mode's contribution to the apsidal motion to differ from its static tidal contribution.

The resonant term in equation (23) can be written as $(0.5 \delta r_\alpha / r_\alpha)/(\delta r_\alpha^2 + d_\alpha^2)$. Thus, for $d_\alpha \gg \delta r_\alpha$, $\zeta_n$ scales as $d_\alpha^{-2}$. For $d_\alpha \ll \delta r_\alpha$, $\zeta_n$ is independent of the damping, $d_\alpha$, and is proportional to $\delta r_\alpha^{-1}$. In addition, for $\delta r_\alpha < 0$, $\zeta_n$ can be less than zero, which corresponds to apsidal recession, a result opposite to that expected from the static tidal formula. The dependence of $\zeta_n$ on $\delta r_\alpha$ is shown in figure (2) for several values of $b_\alpha$; $\Omega_*$ is taken to be zero and $d_\alpha = 0.0005 \lesssim \delta r_\alpha$. It is clear that $\zeta_n$ scales as $\delta r_\alpha^{-1}$ for small $\delta r_\alpha$. However, for larger $\delta r_\alpha$, when the off-resonance terms in the Fourier series make an increasingly dominant contribution to the periastron advance, $\zeta_n$ falls off more slowly. For $b_\alpha \gtrsim 8$, $\zeta_n$ does not depart much from zero because of the small value of the resonant term that is limited by a finite damping ($d_\alpha$) as well as the lower limit to $|\delta r_\alpha|$ imposed by orbital evolution.

The effect of a resonance associated with $m = \pm 2$ modes on $\zeta_n$ depends on the stellar rotation rate, whereas $m = 0$ resonances are, to first order, unaffected by stellar rotation. From equations (23)–(25) we see that, for $b_\alpha \gtrsim 2\Omega_*/\Omega_p$, the effect of stellar rotation is to increase the resonance frequency by $2s$, and thus the effective $b_\alpha$ value of the mode, near resonance, by about $2s\Omega_o/\Omega_p = 2\Omega_*/\Omega_p$.

For $b_\alpha \lesssim 2\Omega_*/\Omega_p$ the dependence of $\zeta_n$ on the stellar rotation rate is more complicated due to an additional possible resonance at the nearest integer to $2s - r_\alpha$, for which the effective $b_\alpha$ value of the mode is $(2\Omega_* - \omega_\alpha)/\Omega_p$. Thus if $\Omega_* \lesssim \omega_\alpha < 2\Omega_*$, rotation causes the effective $b_\alpha$ value of the mode to decrease further below 2 causing $|\zeta_n|$ to decrease. However, for $\omega_\alpha \lesssim \Omega_*$, the effect of rotation is to increase $|\zeta_n|$.

For most physical systems, it is highly unlikely to find a low order $p$ or $g$ mode with $b_\alpha \lesssim 2\Omega_*/\Omega_p$ and thus the effect of rotation is to increase $\omega_\alpha$ by about $2\Omega_*$ and consequently to decrease $|\zeta_n|$. However, the quasi-toroidal modes that exist in rotating stars have $\omega_\alpha \approx \Omega_*$, and thus their effective $b_\alpha$ values, near resonance, can be about 2, thus enhancing their contribution to the periastron advance (this is discussed in greater details in §5).

Finally, we note that for a fixed $b_\alpha$, $\zeta_n$ has a relatively weak dependence on the orbital eccentricity, for $e \gtrsim 0.2$. In figure (3) we show $\zeta_n$ as a function of $b_\alpha$ for $e = 0.4, 0.6$, and $0.8$. Note that different values of the eccentricity, for a fixed value of $b_\alpha$, correspond to different orbits with the same periastron passage time. Physically, the weak eccentricity dependence is due to the fact that the perturbation to the gravitational potential due to tides falls off rapidly away from periastron. Thus, for eccentric orbits, most of the contribution to the apsidal motion comes from near periastron and so orbits of different eccentricity but the same periastron passage time have nearly the same $\zeta_n$. The more rapid decrease of $\zeta_n$ with $b_\alpha$ for smaller $e$ arises from the strong eccentricity dependence of the Fourier coefficients for small $e$ (recall that $D_n \propto \exp(-1.3 b_\alpha e^{-0.25})$). For $e \lesssim 0.2$, the dominant resonance is at $r_\alpha \approx b_\alpha \approx 2$, and this case is described well by equation (13) of the previous section.

## 4. Apsidal Motion for polytropic stars

Having discussed in some detail the effect of a resonance on the contribution of an individual mode to the apsidal motion, we now apply this discussion to the realistic scenario of a star with some prescribed normal mode spectrum. In particular, we consider polytropic stars and investigate when to expect significant departures from the classical apsidal formula.

It is useful to express the mode frequencies of polytropic stars in the following dimensionless form

$$\omega_\alpha^{nd} = \omega_\alpha \sqrt{\frac{R_*^3}{GM_*}}. \tag{26}$$

This nondimensional frequency is a function of polytropic index ($n_{pi}$) but is independent of stellar mass. We rewite the parameter $b_\alpha$, defined in equation (15), as follows:

$$b_\alpha \equiv \frac{\omega_\alpha}{\Omega_p} = \omega_\alpha^{nd} \left[ \frac{M_* d_{min}^3}{(1+e) M_T} \right]^{1/2}, \tag{27}$$

where $d_{min} = a(1-e)/R_*$ is the distance between the two stars at periastron in units of the stellar radius.

As was shown in the previous section, the classical apsidal motion formula can be in error when the value of $b_\alpha$ for at least one of the low order modes of the star is less than some small value, $b_c$, which is of order 7. This condition, together with equation (27), implies that the classical apsidal formula is applicable so long as

$$d_{min} \gtrsim \left(\frac{b_c}{\omega_\alpha^{nd}}\right)^{2/3} \left[\frac{(1+e)M_T}{M_*}\right]^{1/3}. \qquad (28)$$

It should be emphasized that $b_\alpha < b_c$ is a necessary, but not sufficient, condition for the breakdown of the classical formula. Consequently, equation (28) gives a sufficient condition for the validity of the classical apsidal formula.

The $f$, $g_1$ and $p_1$ modes of stars generally have the largest overlap integrals ($|Q_{nl}|$), and since $k$ is proportional to $Q_{nl}^2$, they contribute the most to the apsidal constant. Therefore, if a resonance associated with a $g_n$ mode is to cause deviation from the classical apsidal formula, its contribution to the apsidal constant must be greater than approximately $(Q_{02}/Q_{n2})^2$ times its classically expected value.

The overlap integrals decrease rapidly with the number of radial nodes, $n$, as is shown in figure (4) where we plot the overlap integrals, in units of $(R_*^5/G)^{1/2}$, as a function of $n$ for several polytropes. The decrease of $|Q_{n2}|$, for fixed $n$, with $n_{pi}$ does not imply that stars of higher polytropic index are more likely to lead to deviation from the classical apsidal formula since mode frequencies and thus $b_\alpha$ values also increase with $n_{pi}$. The rapid decrease of $|Q_{n2}|$ with $n$ means, however, that deviation from the classical formula is most likely to arise due to a resonance involving the $f$ or low order $g$ modes. Resonances involving higher order $g$ modes cannot modify the classical result much, unless the modes' overlap integrals happen to be within a factor of a few of the $f$-mode.

We now quantify the likelihood of significant error in the classical apsidal motion formula due to resonance involving low order modes. Since the orbital and modal frequencies are independant quantities, $\delta r_\alpha$ should be treated as a random number with a value between -0.5 and 0.5. Thus, the value of $|\delta r_\alpha|$ required to get a certain value of $\zeta_n$ is a good measure of the probability of failure of the classical apsidal formula. We thus define a probability function, $P_n$, which is equal to twice the largest value of $|\delta r_\alpha|$ such that a resonance involving a mode of order $n$ leads to a fractional error of $\zeta_n$. $P_n$ is a function of $b_\alpha$, $d_\alpha$, $s$, and the stellar structure. The dependence of $P_n$ on $\zeta_n$, $b_\alpha$ and $Q_{nl}$ is determined by separating out the resonant term in equation (23) and solving for $\delta r_\alpha$ in the limit that $\delta r_\alpha \lesssim d_\alpha$:

$$P_n \propto \frac{D_{r_\alpha}/b_\alpha}{\left[\zeta_n \sum_i Q_i^2/Q_\alpha^2 + 1\right] - g(b_\alpha)}, \quad (29)$$

where $D_{r_\alpha}$ is the Fourier coefficient of the resonant term in the Fourier series and $g(b_\alpha)$, the sum of the non-resonant contributions to $\zeta_n$, is a generally increasing function of $b_\alpha$. Thus, for modes that make a small contribution to the periastron advance in the static tide limit, i.e., $\sum_i Q_i^2/Q_\alpha^2 \gg 1$, or for large $\zeta_n$, we have a simple scaling law for $P_n$:

$$P_n \propto \frac{D_{r_\alpha} Q_\alpha^2}{b_\alpha \zeta_n \sum_i Q_i^2}. \quad (30)$$

In figure (5), we show the results of calculating $P_n$ numerically for several different modes and polytropic indices. For all of these calculations we have taken $r_\alpha = 15.0, d_\alpha = 0.0005$, and $\Omega_* = 0$. Note that the main features seen in figure (5) are just what we expect from equation (30). For $\Omega_* \neq 0$, the effect is simply to shift the $b_\alpha$ values, as discussed in §3. This is explicitly illustrated in figure (6) which shows $P_n$ as a function of $b_\alpha$ for the $f$ mode of an $n_{pi} = 3$ polytrope for three different rotation rates. In addition, the weak $r_\alpha$ dependence of $P_n$ follows from the dependence of the Fourier coefficients as discussed in the previous section. This is because calculations of $P_n$ for two different values of $r_\alpha$ and eccentricity, for a fixed $b_\alpha$, are related by the ratio of the Fourier coefficients corresponding to their resonant terms (see eq. [30]). Finally, provided that $d_\alpha \lesssim \delta r_{min}$ (as is likely to be the case), $\delta r_{min}$, not $d_\alpha$, is the smallest physical value of $\delta r_\alpha$, and thus $P_n$ is independent of the dissipation rate.

The probability of deviation from the classical formula, $P_n$, due to a resonance involving higher order $g$ and $p$ modes of these polytropes can be obtained from figure (5) by scaling the variables according to equation (30) and using the overlap integrals from figure (4). For example, the probabilities of obtaining $\zeta_n = 0.3$ due to a resonant $g_3$ mode of an $n_{pi} = 2.0$ polytrope can be calculated using equation (30) and figure (4), and we find them to be about 20 times smaller than those corresponding to $\zeta_n = 2.0$ for the $g_1$ mode of an $n_{pi} = 3.0$ polytrope which are shown in part (c) of figure (5).

Finally, we illustrate how to use the above results to explicitly estimate the system parameters for which the apsidal motion formula should be applicable. For example, to be 90% certain that the static tidal result is within 50% of the exact result, i.e. $\zeta_n \lesssim 0.5$, for a $n_{pi}$=1.5 polytropic star, requires that the $b_\alpha$ value of the $f$ mode should be greater than about 5 (see fig. 5d). This means that for $d_{min} \gtrsim 2.2 \left[(1+e)M_T/M_*\right]^{1/3}$, we expect the classical apsidal formula to be valid to the degree specified.

## 5. Effects of rotation modes on apsidal motion

In a rotating star, the presence of the coriolis force provides an additional restoring mechanism for oscillatory motions. Thus there exists a new family of modes, called the rotation modes or r-modes, which have frequencies of order the stellar rotation rate. Because of their small frequencies, r-modes are likely to be resonantly excited by tidal interactions. In this section we examine the effects of r-modes on apsidal motion.

In a non-rotating star, purely toroidal modes with $\nabla \cdot \boldsymbol{\xi}$=0 and $\xi_r$=0 exist as part of the general solution for nonradial oscillations. These modes correspond to horizontal displacements of the fluid associated with which there is no pressure perturbation; therefore, their mode frequencies are zero. Moreover, these modes are not excited by the tidal force since their density perturbations are zero. However, the toroidal modes become non-trivial in a rotating star. Papaloizou and Pringle (1981) and Rocca (1982) have examined in some detail the importance of r-mode resonances in the context of tidal excitation. We closely follow the approach of Provost, Berthomieu, and Rocca (1981) (hereafter, PBR) to calculate the eigenfunctions and frequencies of r-modes in a slowly and uniformly rotating star. We carry out a perturbative expansion to the structure and eigenmode equations of the star using the non-dimensionalized rotation rate $\tilde{\Omega}_*^2 \equiv \Omega_*^2 R_*^3/GM_*$ as a small expansion parameter. We also assume, as did Papaloizou and Pringle and PBR, that the perturbations are adiabatic and that the perturbation to the gravitational potential can be neglected (Cowling approximation). Saio (1982) has shown that both of these approximations are quite reasonable. Our primary interest lies in the coupling of the r-modes to the quadrupole gravitational potential, and for this we need to calculate their overlap integrals. The basic equations are given in the Appendix. In the following, we summarize the salient features of the r-modes, as relevant for our purposes. To lowest non-vanishing order in $\Omega_*$, the eigenfrequencies are given by:

$$\omega_{n\ell m} = \frac{2m}{\ell(\ell+1)}\Omega_*, \tag{31}$$

For $m = 0$, $\omega_\alpha = 0$, and thus there are no axisymmetric r-modes. To this order, the r-mode eigenfunctions correspond precisely to the toroidal modes of a nonrotating star described above; therefore, even modes with $m \neq 0$ are not excited by the tidal force. Hence we must go to the next order. Higher order physical quantities, as it turns out, can be expressed in terms of lower order quantities. The spherical harmonic functions, however, can no longer adequately describe the rotationally distorted star. The next order quantities have an angular dependence described by a superposition of two spherical harmonics: $Y_{\ell+1,m}$, and $Y_{\ell-1,m}$. This means that one has to be careful in calculating the overlap integrals from equation (4) since in its derivation the angular integration has been performed under the assumption that the mode eigenfunctions are well described by $Y_{\ell m}$. Since the eigenfunctions now consist of sums of $Y_{\ell+1,m}$ and $Y_{\ell-1,m}$, it follows from the orthogonality condition of the spherical harmonics that in order to calculate the overlap integrals $Q_{n2,\pm 2}$, where the subscripts $(2, \pm 2)$ now refer to the $(\ell = 2, m = \pm 2)$th term in the expansion of the gravitational potential, we need the $\ell = 3, m = \pm 2$ r-modes (instead of $\ell = 2, m = \pm 2$ for the p and g-modes, which are based on a spherically symmetric star). We also find that to lowest nonvanishing order, $Q_\alpha \propto \Omega_*$. It is also worth noting that the higher order correction to the eigenfrequency varies as $\Omega_*^3$ which, unlike the lowest order term, depends on the order of the mode. We refer the reader to the Appendix for more details.

In previous sections, we have shown that the tidal excitation of a mode is primarily determined by two parameters, $b_\alpha$ and $Q_\alpha$. We first discuss the $b_\alpha$ dependence of the r-modes. Equation (31) shows that for $\ell = 3, m = \pm 2$, the lowest order frequency is given by $\Omega_*/3$. Note that this is independent of the number of radial nodes ($n$) because, to this order, the mode is purely toroidal. Ignoring higher order corrections, we find $b_\alpha = \Omega_*/3\Omega_p$. However, as shown in §3, stellar rotation effectively shifts the $b_\alpha$ value of these modes (near resonance) by $2(\Omega_* - \omega_\alpha)/\Omega_p$ so that $b_\alpha^{eff} = (5/3)\Omega_*/\Omega_p$. For synchronous rotation at periastron, $b_\alpha^{eff} = 5/3$, which is quite favorable for departure from the classical apsidal formula since $m = \pm 2$ modes couple best to the tidal force for $b_\alpha^{eff} \approx 2$. Non-synchronous rotation moves $b_\alpha^{eff}$ away from its optimal value. Thus, $\zeta_n$ decreases exponentially for super-synchronous rotation.

Using equation (A6) in the Appendix, we compute the quadrupole overlap integrals for several polytropes ($n_{pi} = 1$, 2, and 3). The numerical results for the low-order modes are shown in Table (1). Note that the values in the table are given in terms of the nondimensionalized overlap integral $\tilde{Q}_\alpha$ ($Q_{\ell m} = \tilde{Q}_\alpha \sqrt{R_*^5/G}$). Our results show that the r-modes are unlikely to make a significant contribution to the apsidal motion in spite of their optimal $b_\alpha$ values. The reason for this is that their overlap integrals are too small compared to that of the f-mode. ¿From Table (1), we see that the overlap integrals are generally largest for stars of smaller polytropic index. This is to be expected since stars of higher polytropic index corresponds to more centrally condensed stars, and the effects of rotation for these stars are generally smaller. If, for example, $\tilde{\Omega}_* = 0.1$ for a $n_{pi} = 1$ polytrope, the overlap integral for the $n = 0$ r-mode is three orders of magnitude smaller than that of the f-mode. This makes it unlikely to cause significant deviation from the classical apsidal formula. The situation is not much improved for a more rapidly rotating star since the overlap integral increases only linearly with $\Omega_*$. Indeed, even if we extrapolate our results to the break-up rate (i.e. $\tilde{\Omega}_* = 1$), the overlap integral for the lowest order r-mode of an $n_{pi} = 1$ polyrope is still about one order of magnitude smaller than that of the f-mode. Nevertheless, near break-up rotation, we are no longer justified in ignoring higher order corrections, and our results could be significantly modified.

To summarize, we find that the r-modes are unimportant for the apsidal motion of a slowly and uniformly rotating star. The same conclusion is expected to hold for rapid, but non-synchronous, rotation. This is because the overlap integrals are expected to grow only as some power of the rotation rate (even when we include higher order corrections) while $\zeta_n$ decreases exponentially with increasing $b_\alpha$. Hence, the only regime where r-modes might be important is for near break-up, synchronous rotation, but other methods are needed to investigate this regime in detail.

## 6. Application to AS Cam and DI Her

Close binary systems offer a unique opporunity to test stellar structure theory as well as provide tests for various theories of gravity. Two binary systems, AS Camelopardalis and DI Herculis, are well known because of their anomolous apsidal motions (Guinan and Maloney, 1985; Maloney et al. 1989; Khaliullin et al. 1983, 1991) which are, respectively, about 1/3 and 1/7 of the classically expected values. In this section, we investigate whether resonances of stellar modes with the orbit might be responsible for their anomolous behavior.

AS Cam consists of two main sequence stars ($M_1 = 3.3 M_\odot, R_1 = 2.57 R_\odot, M_2 = 2.5 M_\odot, R_2 = 1.9 R_\odot$) in an $e = 0.1695$ orbit with a period of approximately 3.4 days. The damping time of low order modes of main sequence stars is of order $10^3$ to $10^4$ years; therefore, $d_\alpha \approx 10^{-6}$. Main sequence stars in this mass range have typically both convective and radiative zones and thus are not accurately represented by a single polytropic index. As a crude aproximation, however, we take the stars to be modelled as $n_{pi} = 2.5$ polytropes and calculate the mode frequencies and overlap integrals.

In previous sections, we have, for simplicity, assumed that one of the stars was a point mass. This restriction can easily be relaxed by applying our method to each star and ignoring the second order effects. Using an $n_{pi} = 2.5$ polytrope, we determine that the $b_\alpha$ value for the f mode of the first star ($M_1 = 3.3 M_\odot, R_1 = 2.57 R_\odot$) is about 20, whereas the $b_\alpha$ for the f-mode of the second star is 27.4. Because $b_\alpha \gg -\ln(\delta r_{min})$, it is impossible that an f-mode resonance can give rise to the observed apsidal motion. A more favorable value of $b_\alpha$ is obtained for a high order g-mode. For instance, the $g_{15}$ mode of the star has $b_\alpha \approx 2.2$ and $\sum_i Q_i^2/Q_\alpha^2 \approx 10^7$. In order to obtain $\zeta_n = -0.66$, as required to account for the observed apsidal motion, the $n = 15$ mode of an $n_{pi} = 2.5$ polytrope must have $|\delta r_\alpha| \approx 10^{-9}$. This degree of resonance is, aside from being improbable, unphysical, since it is impossible for the mode to have phase coherence over the required $10^9$ orbits, that is, $\delta r_\alpha \ll \delta r_{min}$. Because of their higher frequencies, p-modes are even less likely to be resonantly excited. Thus, we conclude that the anamolous apsidal motion of AS Cam is not the result of resonances involving g, f or p-modes.

The stellar rotation rates for both stars in AS Cam are observed to be $\Omega_* \approx 1.05 \Omega_o \approx 0.73 \Omega_p$ (Maloney et al. 1989). Thus the stars are rotating somewhat sub-synchronously. The r-modes therefore have $b_\alpha^{eff} \approx 1.2$. From Table (1), (interpolating between $n_{pi} = 2$ and $n_{pi} = 3$), we find that the nondimensionalized overlap integral for the $n = 0$ mode (which generally has the largest overlap integral) is $0.04 \tilde{\Omega}_*$. (Recall from the previous section that, to lowest order, $Q_\alpha \propto \Omega_*$.) With the quoted value for $\Omega_*$, $\tilde{\Omega}_* \approx 0.08$ (for star 1), implying that $\sum_i Q_i^2/Q_\alpha^2 \approx 10^6$. Thus, scaling by equation (30), we find $|\delta r_\alpha| \approx 10^{-8}$ to account for the observed apsidal motion via an r-mode resonance. We can therefore also exclude r-mode resonances as a possible explanation for the anamolous apsidal motion of AS Cam.

DI Herculis consists of a primary with $M_1 = 5.1 M_\odot$ and $R_1 = 2.68 R_\odot$ and a secondary with $M_2 = 4.5 M_\odot$ and $R_2 = 2.48 R_\odot$ in an $e = 0.489$ orbit with a period of about 10.55 days. We also use an $n_{pi} = 2.5$ polytrope as a crude model for these stars. Furthermore, since the periastron passage time and $M_*/R_*^3$ are larger for the DI Herculis system than for AS Cam, the $b_\alpha$ values of the modes are slightly larger than those of AS Cam. Since the dominant dependence of $P_n$ is on $b_\alpha$, we again conclude that f, p, and g-mode resonances cannot cause the observed deviations in the apsidal motion. In addition, the stellar rotation rates of both stars are observed to be $\Omega_* \approx 3.5\Omega_o \approx \Omega_p$ (Guinan and Maloney 1985). Thus the two stars are rotating synchronously at periastron, implying $b_\alpha^{eff} = 1.7$ for the r-modes. Nevertheless, for the primary star, $\tilde{\Omega}_* \approx 0.02$, so that the overlap integrals are even smaller than those of AS Cam. Thus, as before, we can safely rule out r-modes as a possible explanation for the anamolous apsidal motion.

We conclude that the anamolous apsidal motion of DI Herculis and AS Cam is not the result of the breakdown of the static tidal approximation. Other possible explanations, such as the presence of a third body or an inclined spin axis with respect to the orbital plane, are discussed in detail in the above references and references contained therein.

## 6. Summary and Discussion

We have investigated the validity of the classical apsidal formula for tidal distortion by calculating the perturbation to the gravitational potential using a Fourier series expansion of the mode amplitude; this expansion is a good approximation so long as the orbital evolution time scale is such that the change in $r_\alpha$ in one orbit is less than $(\delta r_\alpha)^2$ (where $r_\alpha$ is the ratio of the mode frequency to the orbital frequency and $\delta r_\alpha = r_\alpha - n$, where n is the nearest integer to $r_\alpha$).

We find that the classical apsidal motion formula, which assumes static tides, is very accurate when the dimensionless number $b_\alpha$, which is the ratio of the periastron passage time to the mode period, for the low order f, p, and g-modes, is greater than about 7.

However, when $b_\alpha \lesssim 7$ for one of these modes, dynamical effects become important and the periastron advance can differ significantly from that computed using the static tidal formula. The deviation is largest when one of the low order modes is nearly resonant with the orbit, i.e., when the mode period, for a non-rotating star, is some integral fraction of the orbital period. For a near resonant mode of small $b_\alpha$ the deviation from static tidal formula scales as $\delta r_\alpha^{-1}$, so long as the mode damping time divided by the orbital period is greater than $\delta r_\alpha^{-1}$.

The effective $b_\alpha$ value, near resonance, of a mode in a star rotating synchronously at periastron is generally larger than in a non-rotating star by about two; therefore, the static tidal approximation is better for the p, f, and g-modes of a rotating star.

The contribution of a given mode to the apsidal constant is proportional to $Q_{nl}^2$, where $Q_{nl}$, the overlap integral, is usually largest for the $f$ mode and decreases strongly with the number of radial nodes. Thus, deviation from the classical apsidal motion formula is only likely to arise because of a resonant $f$ mode or a resonant low order $g$ mode whose overlap integral is within a factor of a few of the $f$ mode overlap integral. We have quantified the probability of deviation from the classical apsidal formula by calculating how close the frequency of a low order mode must be to a resonance, $\delta r_\alpha$, in order to obtain a certain deviation from the classical formula as a function of $b_\alpha$ for several polytropes (§4).

For rotating stars there exists a new family of modes (r-modes), where the restoring force is the coriolis force, which have frequencies of order the stellar rotation rate. These modes have optimal effective $b_\alpha$ values and can easily be resonantly excited by the tidal force. We have investigated their effect on apsidal motion for slowly and uniformly rotating stars and find that their overlap integrals are generally too small for them to be important. Only when the star is simultaneously rotating near break-up and synchronous at periastron could the r-modes possibly become important.

We have applied our dynamical calculations of the periastron advance to AS Cam and DI Herculis in order to determine if the discrepancy between the observed and predicted apsidal motion for these systems is due to the break down of static tidal approximation for these systems. We conclude that this is very unlikely to be the case. Given the observational failure to find a third body in either system, we suspect that the anomalous apsidal motion for these systems is likely to be caused by an inclined spin-orbit coupling, as has been suggested by Shakura (1985) and Company et al. (1988).

**Acknowledgment:** We thank Scott Tremaine and John Moffat for suggesting to look at the abnormal apsidal motion of DI Herculis. This work was completed while PK was visiting the Institute for Advanced Study, Princeton. He is very grateful to John Bahcall for his support and hospitality at the Institute. This work was funded in part by a NASA grant NAGW-3936.

# Appendix

In this Appendix, we summarize the basic equations for r-modes in a slowly and uniformly rotating star as described in PBR. We also give the expression for the overlap integral.

Express all physical quantities in the natural units where $M_* = R_* = G = 1$. Let $\Omega_*$ be the rotation rate of the star and let the z-axis be the axis of rotation. We work in the rotating frame of the star. Since $\Omega_*$ is assumed to be small, we can express the equilibrium structure of the star as

$$r = x(1 - \Omega_*^2 \alpha(x) \cos^2 \theta), \tag{A1}$$

where $x$ labels the equipotential surfaces, and $\alpha(x)$ describes the rotational distortion to lowest order. We calculate $\alpha(x)$ numerically using the Clairaut-Legendre expansion as discussed in Tassoul (1978). Similarly, we can expand the eigenfrequencies and eigenfunctions in powers of $\Omega_*^2$ as follows:

$$\omega = \Omega_* \omega_0 (1 + \Omega_*^2 \omega_1)$$
$$\xi_r = \Omega_*^2 \xi_r^1 e^{im\phi}$$
$$\xi_\theta = (\xi_\theta^0 + \Omega_*^2 \xi_\theta^1) e^{im\phi}$$
$$\xi_\phi = (\xi_\phi^0 + \Omega_*^2 \xi_\phi^1) e^{im\phi}. \tag{A2}$$

The zeroth order fluid equations correspond to the conservation of mass and angular momentum about the z-axis. Its solutions are

$$\omega_0 = \frac{2m}{\ell(\ell+1)}$$
$$\boldsymbol{\xi}_{n\ell m}^0 = \left[0, im C_{\ell m}(x) \frac{P_\ell^m(\cos\theta)}{\sin\theta}, -C_{\ell m}(x) \frac{d}{d\theta} P_\ell^m(\cos\theta)\right] e^{im\phi}, \tag{A3}$$

where $C_{\ell m}(x)$ describes the mode structure in the radial direction (unrelated to the Fourier coefficients defined earlier!). To zeroth order, $C_{\ell m}(x)$ is completely arbitrary (which explains why the zeroth order frequency is independent of the order of the mode); it is determined only through the first order equations. PBR derived the following second order ODE for $C_{\ell m}(x)$:

$$\frac{d^2}{dx^2}C_{\ell m} + \frac{d}{dx}\ln\left(\frac{\rho x^4}{Ag}\right)\frac{dC_{\ell m}}{dx} + \left[\frac{Ag}{x^2}\lambda_1\omega_1 - \frac{\ell(\ell+1)}{x^2} + A\frac{d}{dx}\ln\left(\frac{x^4}{g}\right)\right.$$
$$\left. + \frac{\lambda_2}{x}\frac{d}{dx}\ln\left(\frac{x^2}{Ag\rho}\right) - \frac{Ag}{x^2}(\lambda_3\alpha(x) + \lambda_4 x\frac{d}{dx}\alpha(x))\right]C_{\ell m} = 0, \quad (A4)$$

where $\rho, P, g$, and $\gamma$ have their customary meanings, and

$$A \equiv \frac{1}{\rho}\frac{d\rho}{dx} - \frac{1}{\gamma P}\frac{dP}{dx}.$$

The $\lambda$s depend only on $\ell$ and $|m|$ and are defined as:

$$\lambda_1 = 0.25 D_\ell \ell^3 (\ell+1)^3$$
$$\lambda_2 = D_\ell[(\ell+1)^5 K_-(\ell-1)K_+(\ell) - \ell^5 K_-(\ell)K_+(\ell+1)]$$
$$\lambda_3 = 0.5 D_\ell \ell^2 (\ell+1)^2 m^2$$
$$\lambda_4 = 0.5 D_\ell \ell^2 (\ell+1)^2 [(\ell+1)^2 K_-(\ell-1)K_+(\ell) + \ell^2 K_-(\ell)K_+(\ell+1)]$$

with

$$K_+(\ell) = \frac{\ell + |m|}{2\ell + 1}$$
$$K_-(\ell) = \frac{\ell - |m| + 1}{2\ell + 1}$$
$$D_\ell = [(\ell+1)^4 K_-(\ell-1)K_+(\ell) + \ell^4 K_-(\ell)K_+(\ell+1)]^{-1}$$

It is clear from equation (A4) that $C_{\ell m}$ is independent of the sign of $m$. Having obtained $C_{\ell m}$, one can then go back to the first order fluid equations and solve for the first order quantities in terms of $C_{\ell m}$. Consequently, the overlap integral, which can be expressed in terms of $C_{\ell m}$ and its derivative, depends only on $|m|$. Hence, for notational simplicity, we assume that $m > 0$ in subsequent discussions. Since $C_{\ell m}$ gives the horizontal displacement, it must be zero at the center and finite at the surface. Equation (A4), together with these boundary conditions, constitute a well-defined eigenvalue problem with eigenvalue $\omega_1$. We have numerically computed $C_{32}(x)$ and the corresponding $\omega_1$ for several polytropes since these are the modes that couple to the tidal force (see below). The results for $\omega_1$ are shown in Table (2). Note that when the correction frequency $\omega_1\Omega_*^2$ becomes comparable to 1, equation (A4) is no longer valid. Since $|\omega_1|$ tends to increase with mode order $n$, for a given $\Omega_*$, there are only a limited number of low to moderate order modes we can obtain from the PBR procedure.

To calculate the overlap integrals, we start with a definition more general than equation (4):

$$Q_{n\ell m} \equiv -\sum_{\ell' m'} \int x^4 dx d\Omega \nabla \cdot (\rho \boldsymbol{\xi}_{n\ell' m'}) Y^*_{\ell m}(\Omega), \qquad (A5)$$

where the factor of $Y^*_{\ell m}$ comes from the standard expansion of the gravitation potential $1/|\mathbf{R}-\mathbf{r}|$ in terms of spherical harmonics. It is easy to see that the above definition reduces to equation (4) when the displacement eigenfunction involves only one spherical harmonic. From the above equation, we can readily calculate the overlap integrals in terms of $C_{\ell m}$ and obtain, to lowest non-vanishing order,

$$|\tilde{Q}_{\ell m}| = 4m\Omega_*^2 \sqrt{\frac{4\pi}{2\ell+1}} \sqrt{\frac{(\ell+m)!}{(\ell-m)!}} \left| \frac{(\ell+m+1)}{(2\ell+3)(\ell+1)^2} B_+ - \frac{(\ell-m)}{(2\ell-1)\ell^2} B_- \right|, \qquad (A6)$$

where

$$B_+ = \int dx \frac{x^4}{g} \rho C_{\ell+1,m}(x) \left[ \ell - 3 + x \frac{d\ln g}{dx} \right]$$

$$B_- = \int dx \frac{x^4}{g} \rho C_{\ell-1,m}(x) \left[ \ell + 4 - x \frac{d\ln g}{dx} \right].$$

We normalize the eigenfunctions such that each mode carries unit energy. Hence $C_{\ell m} \propto 1/\Omega_*$ and $Q_\alpha \propto \Omega_*$. Note that in general, to calculate the overlap integral corresponding to the $(\ell, m)$th term in the expansion of the gravitational potential, we need the r-mode eigenfunctions labelled by $(\ell+1, m)$ and $(\ell-1, m)$. For tidal excitation, we are mainly interested in quadrupole modes with $\ell = 2, m = 2$, in which case we need to calculate $C_{32}(x)$ to obtain the required overlap integrals.

**Figure Captions**

FIG. 1.—A comparison of $\eta_n$, the fractional deviation, for a given mode, from the periastron advance predicted by the classical apsidal formula, with $\zeta_n$, the fractional deviation, for a given mode, from the perturbation to the gravitational potential at periastron predicted by the classical apsidal formula. Part (a) of this figure shows $\zeta_n$ and $\eta_n$ as functions of $b_\alpha$ while Part (b) compares the two as functions of $\delta r_\alpha$. Since most of the contribution to the periastron advance comes from the perturbation to the gravitational potential near periastron, $\zeta_n$, as illustrated by this figure, is a reasonable approximation to $\eta_n$; consequently, in this paper, we adopt $\zeta_n$ as an approximate measure of the deviation from the classical apsidal formula.

FIG. 2.—The fractional deviation ($\zeta_n$), for a given mode, from the classical apsidal formula as a function of $\delta r_\alpha$ for different values of $b_\alpha$. $d_\alpha = 0.0005 \lesssim \delta r_\alpha$ and thus $\zeta_n \propto \delta r_\alpha^{-1}$. This dependence, however, sets in at increasingly larger values of $\delta r_\alpha$ for increasing values of $b_\alpha$ as the non-resonant terms in $\zeta_n$ make an increasingly dominant contribution to the periastron advance. By $b_\alpha \approx 8$, the $\delta r_\alpha^{-1}$ dependence only sets in at $\delta r_\alpha \approx d_\alpha$ and thus the resonance has little effect. We therefore emphasize that resonances effect $\zeta_n$ and deviation from the classical apsidal formula is only expected for small $b_\alpha$.

FIG. 3.—$\zeta_n$ as a function of $b_\alpha$ for three values of the orbital eccentricity: $e = 0.4, 0.6$, and $0.8$. Keeping $e$ fixed and varying $b_\alpha$ corresponds to varying $r_\alpha$. For a fixed value of $b_\alpha$, the deviation from the classical apsidal formula is approximately the same for orbits of different eccentricities. Physically, this is due to the fact that the tidal potential falls off rapidly from periastron and thus the coupling of a given mode to the tidal force is primarily determined by the periastron passage time.

FIG. 4.—The overlap integral, $Q_{n2}$, in units of $(R_*^5/G)^{1/2}$, as a funtion of the number of radial nodes, $n$, for stars of different polytropic index, $n_{pi}$. $n < 0$ is for $g$ modes and $n > 0$ is for $p$ modes. The value of the apsidal constant is proportional to $Q_{n2}^2$ and thus the $f$ mode and low order $p$ and $g$ modes contribute the most to the apsidal constant. The rapid decrease of $Q_{n2}$ with $n$ therefore implies that high $n$ $g$ modes, although they are more likely to have $b \approx 1$, are unlikely to cause deviation from the classical apsidal formula because of their small overlap integrals.

FIG. 5.—$P_n$, defined to be twice the largest value of $\delta r_\alpha$ such that a certain value of $\zeta_n$ is obtained as a function of $b_\alpha$, $d_\alpha$, $s$, etc., is a good measure of the probability of failure of the classical apsidal formula. Parts (a), (b), and (c) of this figure show $P_n$ for several values of $\zeta_n$ as a function of $b_\alpha$ for the $f$, $g_1$, and $p_1$ modes of an $n_{pi} = 3.0$ polytrope, respectively. Part (d) shows $P_n$ for the $f$ mode of an $n_{pi} = 1.5$ polytrope. We find, however, that the $f$ modes of polytropes with indices between 1.5 and 2 exhibit the same deviation from the classical apsidal formula so these results are also valid for $n_{pi} = 2$. $P_n$ for additional $p$ and $g$ modes of these polytropes and for additional $\zeta_n$ can be obtained from the results in this figure by a simple scaling law given in equation (30).

FIG. 6.—$P_n$, for $\zeta_n = 0.5$, as a function of $b_\alpha$ for three different rotation rates: $\Omega_* = 0.0$, $\Omega_* = 0.25\Omega_p$, and $\Omega_* = 0.5\Omega_p$, where $\Omega_p$ is the angular velocity of the star at periastron. For modes with $b_\alpha \gtrsim 1$, rotation effectively increases the $b_\alpha$ value of the mode by approximately $2\Omega_*/\Omega_p$, thus making the static tidal approximation more applicable.